\documentclass[%
 reprint,
superscriptaddress,
 amsmath,amssymb,
 aps,prx,
]{revtex4-2}

\usepackage{graphicx}
\usepackage{dcolumn}
\usepackage{bm}
\usepackage[hidelinks]{hyperref}
\hypersetup{
    colorlinks=true,
    linkcolor=blue,
    filecolor=blue,
    urlcolor=blue,
    citecolor = blue,
    pdftitle={State-selective EIT for quantum error correction in neutral atom quantum computer}
    }
\usepackage{physics}
\usepackage{xfrac}
\usepackage[all]{hypcap}

\begin{document}

\preprint{}

\title{State-selective EIT for quantum error correction in neutral atom quantum computers}

\author{Felipe Giraldo}
    \affiliation{Physics Department, Pennsylvania State University, University Park, Pennsylvania 16802, USA}
\author{Aishwarya Kumar}
    \altaffiliation[Present address: ]{James Franck Institute and Department of Physics, University of Chicago, Chicago, Illinois 60637, USA}
    \affiliation{Physics Department, Pennsylvania State University, University Park, Pennsylvania 16802, USA}
\author{Tsung-Yao Wu}
    \altaffiliation[Present address: ]{Atom Computing, Inc., Berkeley, California 94710, USA}
    \affiliation{Physics Department, Pennsylvania State University, University Park, Pennsylvania 16802, USA}
\author{Peng Du}
    \affiliation{Physics Department, Pennsylvania State University, University Park, Pennsylvania 16802, USA}
\author{David S. Weiss}
    \email{dsweiss@phys.psu.edu}
    \affiliation{Physics Department, Pennsylvania State University, University Park, Pennsylvania 16802, USA}

\date{\today}

\begin{abstract}
We propose a way to measure the qubit state of an arbitrary sub-ensemble of atoms in an array without significantly disturbing the quantum information in the unmeasured atoms. The idea is to first site-selectively transfer atoms out of the qubit basis so that one of the two states at a time is put into an auxiliary state. Electromagnetically induced transparency (EIT) light will then protect most states while detection light is scattered from atoms in the auxiliary state, which is made immune to the EIT protection by angular momentum selection rules and carefully chosen light polarization. The two states will be measured in turn, after which it is possible to recool and return the atoms to a qubit state. These measurements can be the basis of quantum error correction.

\end{abstract}

\maketitle


\section{\label{sec:level1}Introduction}

There has been substantial recent progress in developing neutral atom qubits. Along with better control of atom locations and better vibrational cooling in systems with $>$50 atom qubits \cite{Endres2016_sort1D, Barredo2016_sort2D, Barredo2018_sort3D, Kumar2018_sort3D, Mello2019_sort2D100, Young2020_150atom, Ebadi2020_sort256atom, Scholl2020_sort196atom, Jenkins2021-wn}, there have been advances in high fidelity state detection \cite{Wu2019_SG, Madjarov2020_Sr2Q}, single-qubit gate fidelity \cite{Xia2015_1Q, Wang16_1Q, Ma2021-zq, Sheng18-g1Q} and two-qubit gate fidelity \cite{Levine2018_Rb2Q, Graham2019_Cs2Q, Levine2019_Rb2Q, Madjarov2020_Sr2Q, Schine2021-em, Graham2021-yp, Yang2020}. Continued progress is necessary on all these technical fronts, but it is not too early to think seriously about the remaining element needed for universal quantum computation: the ability to correct errors \cite{Shor1995_QEC, Steane1996_QEC, Gottesman2009_QEC}. Quantum error correction (QEC) requires being able to measure the states of selected qubits while preserving the quantum information in the rest. One-way quantum computation, where entanglement is initially encoded into the system, also requires site-specific measurements \cite{Raussendort2001_1wayQC}. The challenge on this front for neutral atoms, as well as for trapped ions, is that their qubit states are typically detected by scattering photons, which can be rescattered by surrounding “spectator” atoms, randomly changing their quantum states. The challenge is greater when the atoms are more closely spaced and it is especially large in 3D arrays, where there are typically atoms in the path of all scattered photons.

A few ideas have been previously suggested for meeting the selective measurement challenge. Two atomic species can be used, one for computation and one for measurement \cite{Beterov2015_2species,dualspecies1,dualspecies2}. In ion systems, the quantum logic clock is based on this concept and has yielded very precise state detection \cite{Schmidt2005_ionQlogic, Hume2007_ionQHighF}. But it has not been implemented in systems with many qubits or in neutral atom systems. Alkaline earth-like atoms present the possibility of storing qubit states in metastable triplet levels that do not rescatter light from singlet scattering transitions. These atoms could obviate the need for the type of spectator protection proposed here, but perhaps at the cost of extra, routine manipulation of the states of spectator atoms, which would come with its own error costs. Another viable approach that can be applied in some trapping configurations, is to move the atoms to be measured far enough away from the other atoms to sufficiently suppress rescattering when their states are being detected \cite{LukinMovement,LukinQECRydberg}. 

Spatially selective measurement of an alkali atom has been demonstrated using EIT to suppress repumping from all but well-localized spatial positions \cite{Miles2013, Subhankar2019, McDonald2019_superresolution}. Such an approach has not been adapted for qubit state measurements. The idea that we propose here combines site-selective state transfers and EIT to yield high fidelity localized state detection while minimally affecting surrounding quantum information. Such detection can be lossless and allows for recooling the measured qubits.

The general approach we propose starts with identifying a sub-sample of atoms to be measured and site-selectively mapping the qubit states of those atoms onto two auxiliary states. Then the two auxiliary states will be detected in turn in the presence of EIT light that suppresses light scattering from all other states, including the qubit basis. Having thus performed a state measurement on the selected atoms, they can be transferred back to one of the qubit basis states.

In the next section we will expand on this overview and the need for protection. Section~\ref{SSEIT} delves into the core part of the procedure, the implementation of state-selective EIT. Section~\ref{Methodology} describes how we numerically analyze the full model. In Section~\ref{Scheme1Section}, we characterize the performance of an implementation that involves a cycling transition, which we call Scheme 1. Our analysis takes into account all relevant hyperfine states. In Section~\ref{Scheme2Section}, we present a similar analysis for an implementation without a cycling transition, Scheme 2. The lack of a cycling transition for detection adds some complications, but also allows atoms to be cooled and reused. Finally,in Section~\ref{PolarizationImperfectionsSection}, we consider the effects of polarization imperfections and how they will impact real experiments and thereby demonstrate the practical feasibility of these proposals.

\section{Overview}

\begin{figure}
\includegraphics[width=0.45\textwidth]{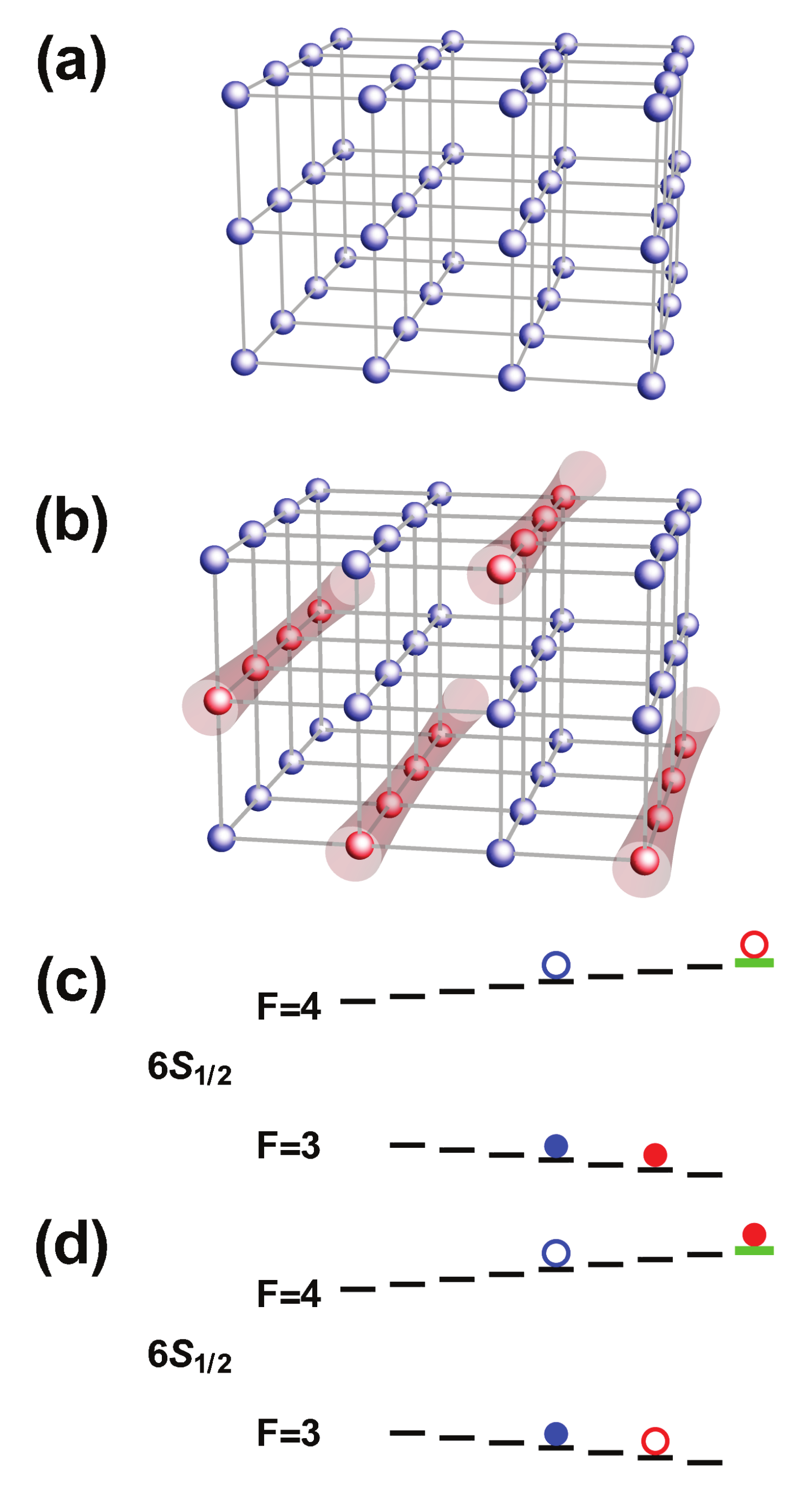}
\caption{\label{overview}Overview of the measurement. (a) Initially, atoms in a 3D optical lattice are in random superpositions in the qubit basis (shown in blue). (b) A select group of atoms are transferred to states out of the qubit basis (shown in red). (c) The qubit and non-qbuit bases in the Cs ground state. The empty (filled) circle blue states are transferred to the empty (filled) circle red states using a series of stimulated pulses. The red state population in the green sublevel is detected, while all other states are EIT protected. (d) Another series of stimulated pulses exchanges the empty and filled circle red states, allowing the other state population to be detected. }
\end{figure}

The proposed measurement starts with an array of atomic qubits in arbitrary states, a 3D version of which is shown in Fig.~\hyperref[{overview}]{\ref*{overview}(a)}, where the blue spheres represent atoms that can have any superposition in the qubit basis. Stage 1 of our approach will be to coherently map the internal states of selected atoms, illustrated by the red spheres in Fig.~\hyperref[{overview}]{\ref*{overview}(b)}, out of the qubit basis into a basis that includes a stretched state. Fig.~\hyperref[{overview}]{\ref*{overview}(c)} shows an example, for Cs atoms, of possible states to use for both the qubit basis and the new basis. To reach the new basis, the atoms to be measured will first be transferred out of the qubit basis using site-specific addressing \cite{Wang2015_3Daddr}, either in series or in parallel. Once out of the qubit basis, one qubit state will be mapped onto the stretched state, and the other qubit state will be mapped onto an intermediate state, as shown in Fig.~\hyperref[{overview}]{\ref*{overview}(c)} (see Appendix~\ref{Mapping} for a particular example of a pulse sequence for making this mapping). Since the intent is to measure the probability of being in each initial state, it is not necessary to maintain the coherence between the two states during these transfers.

Stage 2 is to detect atoms in and only in the stretched state. The central idea is to use EIT to suppress the scattering of detection light from all occupied magnetic sublevels in the upper hyperfine level except the stretched state (and sometimes the adjacent sublevel), as illustrated in Fig.~\hyperref[{overview}]{\ref*{overview}(c)}. The lower hyperfine states will be protected by their detuning. There are several ways to accomplish this state-selective EIT protection, which we will discuss in the next section.

Stage 3 consists of using stimulated pulses, either microwave or Raman, to exchange the magnetic sublevels into which the selected atoms were originally placed, as shown in Fig.~\hyperref[{overview}]{\ref*{overview}(d)}. The EIT-protected detection from the stretched state can then be repeated. Detecting both states allows atom loss to be monitored.

Finally, Stage 4 consists of a reversal of Stage 1, returning each of the measured atoms to one of the qubit states. The particular state each atom is returned to will depend, in a known way, on the result of the measurement. As in Stage 1, the transitions need only be made site-specific for the final transfer.

Before delving into the details of our proposed protection scheme, we will elaborate on the need for such protection and its requirements. Direct illumination of non-target atoms causes too large an error, even with EIT protection, so it must be minimized. The main concern is rescattering of the light emitted by imaged atoms, although there may also be a small amount of stray detection light due to imperfect beams or scattering from surfaces. For resonant light, the reabsorption cross section is given by \cite{Loudon}:

\begin{equation} \label{Errorequation3}
    \sigma = \frac {\lambda^2}{2 \pi}
\end{equation}

The probability of rescattering resonant light by a single Cs atom 5 $\mu$m away is 0.0004 per photon. The number of photons needed for detection, $\gamma$, depends on the detection scheme. Using resonant cavities \cite{CavityDetection1,CavityDetection2} or phase-sensitive imaging techniques \cite{PhaseContrast1,PhaseContrast2} allows atom detection with tens of scattered photons, while more traditional approaches need at least several hundred scattered photons \cite{photodetection1,photodetection2,photodetection3,photodetection4}. Here we will assume that atom detection can be achieved with 100 scattered photons; the expected error in any particular experiment can be rescaled accordingly. For $\gamma$=100, the error rate on the adjacent atom is 0.04, obviously too high. 

Since the spectator atom is not illuminated with the detection light, rescattering can be decreased by increasing the detuning of the detection light. However, the reduction is proportional to the required detection light intensity, making beam imperfection and surface scattering a proportionally worse problem. Furthermore, off-resonant detection leads to more hyperfine changing spontaneous emission, compromising most state detection schemes. So generally, and by a large margin, protection against rescattering is needed.

A complete analysis of rescattering in an array would have to take into account details like the specific geometry of the array, the position of the target atoms inside the array and the polarization of the imaging light. To simplify the problem, we will assume that the imaged atom is in the middle of the array and emits light spherically symmetrically. We will limit the discussion to large enough 2D and 3D arrays that we can ignore the specific lattice configuration and approximate the surrounding distribution of atoms as shells.

It is useful to define a target maximum total error caused by imaging a single atom, $\mathcal{E}$, which can be $\mathcal{E}<10^{-4}$, a conservative error rate for quantum error correction \cite{Gottesman2009_QEC}. If secondary rescattering can be ignored, then the total error caused by rescattering can be approximated by the sum of the errors at all potentially affected atoms, which is given by:

\begin{equation} \label{Errorequation1}
    \mathcal{E}= \sum_{i\in spectators} p_i \, \gamma
\end{equation}

where $p_i$ is the probability that a spectator atom $i$ rescatters a photon from the imaged atom. Assuming spherical light emission, the equation can be recast as:

\begin{equation} \label{Errorequation2}
    \mathcal{E}= \sum_{shells} \frac {\sigma}{4 \pi r_i^2} \, N_i \, \gamma
\end{equation}

where $r_i$ defines a shell of atoms at a given radius around the imaged atom and $N_i$ is the number of spectator atoms in this shell. 

We will characterize the effect of the EIT light with the factor $R$, which is the suppressed scattering rate normalized to the scattering rate with no protection. Thus $R$ is the suppression factor for the reabsorption of scattered photons, which will be the main figure of merit throughout the paper. The total error becomes:

\begin{equation} \label{Errorequation4}
    \mathcal{E} = \sum_{shells} \frac {\lambda^2}{8 \pi^2 r_i^2} \, R \, N_i \, \gamma
\end{equation}

The number of atoms in a given shell depends on the dimension of the array. For 2D arrays with lattice spacing $L$, the number of atoms in a given radius is approximately:

\begin{equation} \label{Errorequation5}
    N_{i,2D} = \frac {2 \pi r_i}{L}
\end{equation}

so the total error in 2D is given by: 

\begin{equation} \label{Errorin2Dequation}
    \mathcal{E}_{2D}= \sum_{shells} \frac {\lambda^2}{4 \pi r_i L} \, R \, \gamma
\end{equation}

which grows with array size, but very slowly. A given maximum allowed error dictates the maximum size the array can have. 

3D arrays offer benefits for quantum computation, such as enhanced connectivity and favorable scalability \cite{scalability}. The number of atoms in a shell in 3D is given by: 

\begin{equation} \label{Errorequation6}
    N_{i,3D} = \frac {4 \pi r_i^2}{L^2}
\end{equation}

so the error is given by:

\begin{equation} \label{Errorequation7}
    \mathcal{E}_{3D}= \sum_{shells} \frac {\lambda^2}{2 \pi L^2} \, R \, \gamma.
\end{equation}

Each new shell adds the same amount of error, so that the expression can be simplified further:

\begin{equation} \label{Errorin3Dequation}
    \mathcal{E}_{3D}= \frac {\lambda^2}{2 \pi L^2} \, R \, \gamma \, n_{shells}
\end{equation}

Imaging single atoms in a 3D array is not possible without direct illumination of spectator atoms, so columns of atoms need to be addressed at a time. Because there is no way for light to exit without passing through spectator atoms, rescattering is more of a problem than in 2D, making EIT protection even more necessary.

For Cs, scattering from the hyperfine state that is not resonant with the imaging light is typically 3 to 4 orders of magnitude lower than scattering from even the EIT-protected resonant state. Therefore, the error rates due to rescattering could be reduced by temporarily transferring all the spectator atoms from the typical qubit basis to a basis with states only in the off-resonant hyperfine manifold. However, this would come at the expense of errors in the transfer and also errors accumulated in this new basis, which would usually be more sensitive to magnetic field noise. We will not further consider this approach in this paper, but simply note it as a possibility.

\section{State-selective EIT protection}\label{SSEIT}

There are many ways to implement EIT protection for this purpose, which can generally be categorized into two types, those with and without cycling transitions for detection. We will present detailed examples of a scheme of each type. In our Scheme 1, the unprotected transition used for detection is a cycling transition. Our Scheme 2 lacks a cycling transition for detection and so requires the use of extra stimulated pulses to avoid dark states. These extra stimulated pulses allow for a natural way to incorporate cooling.
 
We consider ladder-type EIT systems in Cs. The general configuration is as follows: a detection beam is resonant with the transition between a ground state ($6S_{1/2}, \, F=4$) and an intermediate state, which can be either $6P_{3/2},\,F=5$ or $6P_{1/2},\,F=4$. An EIT protection beam is resonant with the transition between the intermediate state and an excited state, which can be either $7S_{1/2},\,F=3$ or $4$. We will refer to the hyperfine states as $\ket{F,m_F}$, with no prime for the ground state (e.g. $\ket{4,4}$), a single prime for the intermediate states (e.g. $\ket{4',4'}$) and double prime for the excited state (e.g. $\ket{4'',4''}$). Our calculations are for particular electronic levels in Cs atoms, but we expect the ideas to carry over to other states and other systems.
 
\begin{figure}
\includegraphics[width=0.45\textwidth]{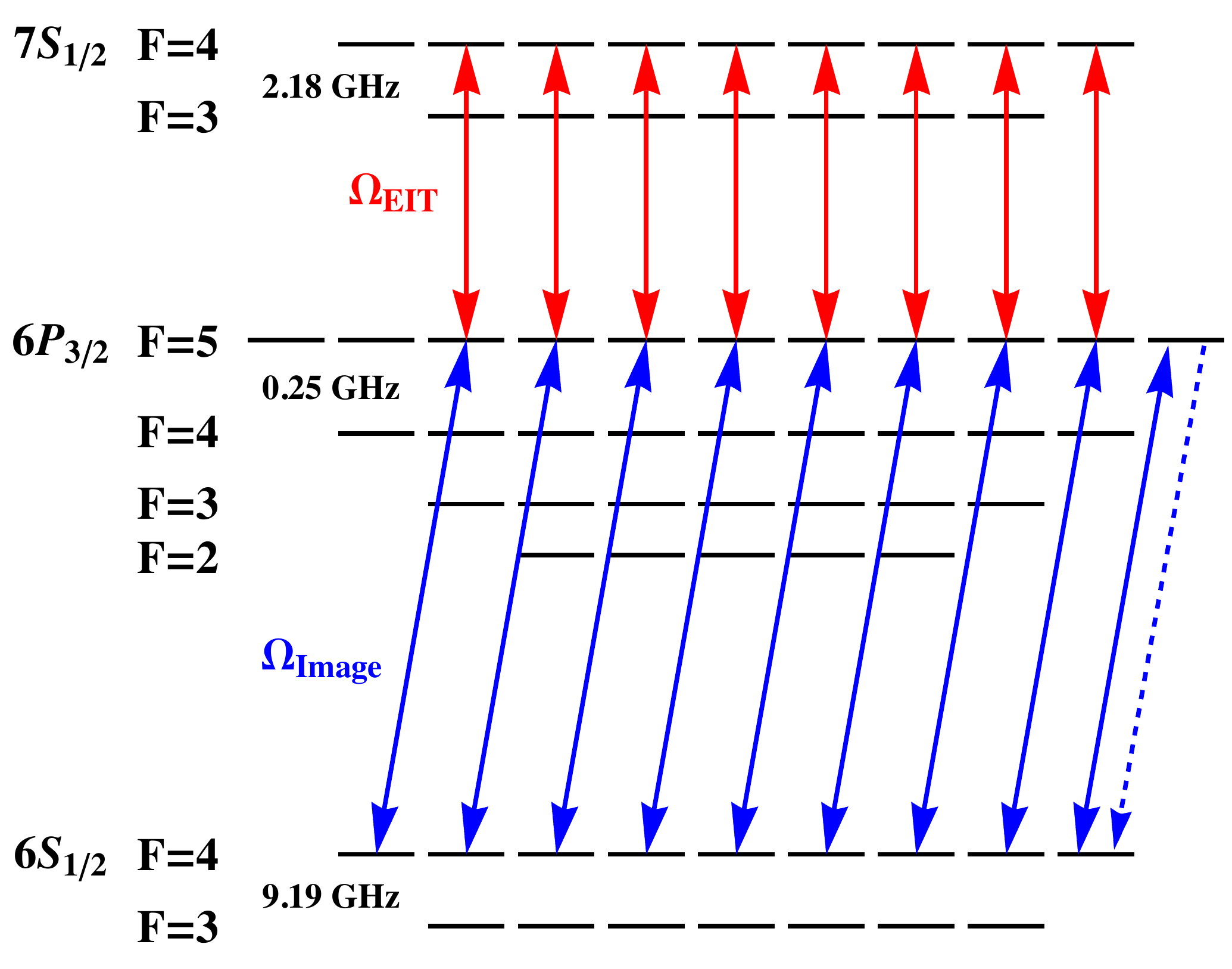}
\caption{\label{scheme1}Scheme 1 EIT-protected detection on the D2 line. The imaging light uses $\sigma^+$-polarization and the EIT protection beam uses $\pi$-polarization. With this configuration all F=4 sublevels are EIT-protected except $\ket{4,4}$, which can be imaged on the cycling transition $\ket{4,4}\longrightarrow\ket{5',5'}$. }
\end{figure}

Scheme 1 is shown in Fig.~\ref{scheme1}. The transition $\ket{4,4}$ to $\ket{5',5'}$ is not EIT-protected, so atoms in $\ket{4,4}$ can scatter light. The non-resonant $F'$ and $F''$ levels significantly impact EIT performance, and sorting out their impact constitutes much of this paper. Before addressing the real system, it is useful to start with a stripped down toy Cs model, with perfect polarization and only the $F'=5'$ intermediate state and the $F''=4''$ excited state.

In this toy model there are 5 sublevels of primary interest. The $\ket{4,4}$ and $\ket{5',5'}$ sublevels form a closed cycling transition for detection. The $\ket{4,0}$, $\ket{5',1'}$ and $\ket{4'',1''}$ states form a ladder type 3-level system. The 3-level Hamiltonian can be diagonalized after making the rotating-wave approximation. On two photon resonance, the (nearly) "non-absorbing" energy eigenstate is given by:
 
\begin{equation}\label{eqn1}
    \ket{\psi}_{NA} = \cos\theta \ket{4,0} - \sin\theta\ket{4'',1''}
\end{equation}

where, 

\begin{equation}
    \tan\theta = \frac{\Omega_{Image}}{\Omega_{EIT}}
\end{equation}

\begin{figure}
\includegraphics{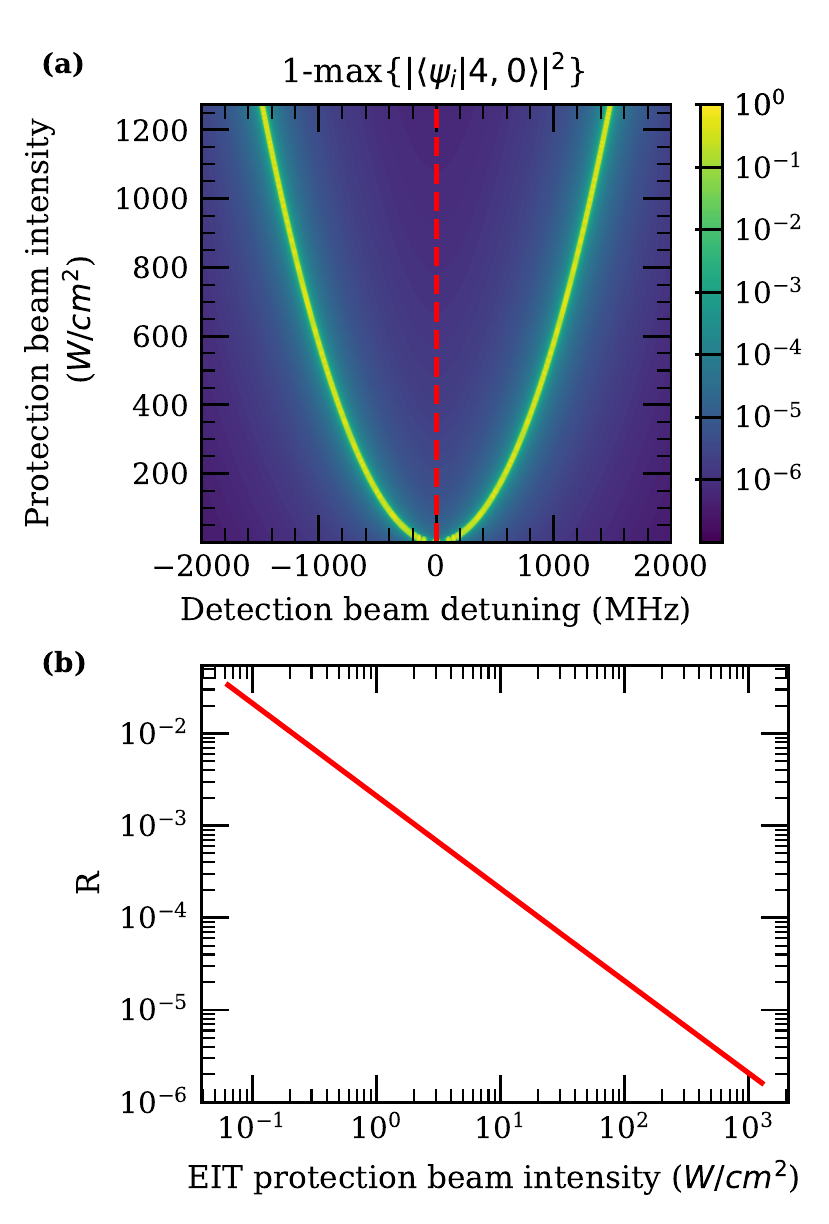}
\caption{\label{3level} EIT-protection of an idealized 3 level-system. (a) The non-ground state population of the EIT-protected qubit state as a function of the detection beam detuning and the protection beam intensity. The red dashed line marks where the detection beam is resonant with the cycling transition. (b) The suppression factor for the reabsorption of scattered photons ($R$) as a function of intensity of the protection beam.}
\end{figure}

Here, $\Omega_{Image}$ is the detection light Rabi frequency and $\Omega_{EIT}$ is the protection Rabi frequency. Light scatters from the non-absorbing state to the extent that it contains the finite lifetime excited state. In the limit, $\Omega_{EIT} >> \Omega_{Image}$, the scattering rate from $\ket{\psi}_{NA}$ decreases linearly with protection beam intensity. Fig.~\hyperref[{3level}]{\ref*{3level}(a)} shows $1-max\{|\bra{\psi_i}\ket{4,0}|^2\}$, the total intermediate and excited state populations for the eigenstate that has the highest projection onto the ground state, as a function of the protection beam intensity (vertical axis) and the detection beam detuning (horizontal axis). Fig.~\hyperref[{3level}]{\ref*{3level}(b)} shows $R$ as a function of EIT protection beam intensity when the detection transition is resonant (along the red line in Fig.~\hyperref[{3level}]{\ref*{3level}(a)}). In the limit of infinite protection intensity, there is perfect EIT protection of $\ket{4,0}$, while detection on the cycling transition is unaffected.
 
In any real atom, the presence of other intermediate and excited levels like hyperfine levels prevents this ideal limit from being approached. The rest of this paper is devoted to analyzing in detail the full model for Cs.

\section{Methodology}\label{Methodology}
We use the QuTiP Python package \cite{Johansson2013_qutip} to solve the master equation and extract populations in various magnetic hyperfine sublevels. For transition frequencies, energies and matrix elements, we adapted the ARC package \cite{ARC} to include the hyperfine splittings. The atoms are initially prepared in different sublevels depending on whether we want to characterize protection or detection. The simulation includes the full hyperfine structure of the ground, intermediate and excited states in the presence of the two light fields (detection and protection beams). The Hamiltonian can be made time-independent by using unitary transformations similar to the well known 3-level EIT scheme and the rotating wave approximation. Only one of the $P$ fine structure manifolds is considered at a time. Since this is not enough to account for the entire spontaneous emission from the $7S_{1/2}$ state (which requires the other $6P$ state), the spontaneous emission from $7S_{1/2}$ to the simulated $6P$ state is scaled up to match the experimentally known lifetime of the $7S_{1/2}$ state. For all simulations, we keep the detection beam intensity fixed and vary the protection beam intensity. We used a beam of $12.7\:{\mu}\text{W}/\text{cm}^2$ for the detection beam. This is an order of magnitude below the saturation intensity, so all our results for scattering rates scale linearly in this regime.

\section{\label{Scheme1Section}EIT protection Scheme 1}

\begin{figure*}[htb!]
\centering
\includegraphics[width=\textwidth]{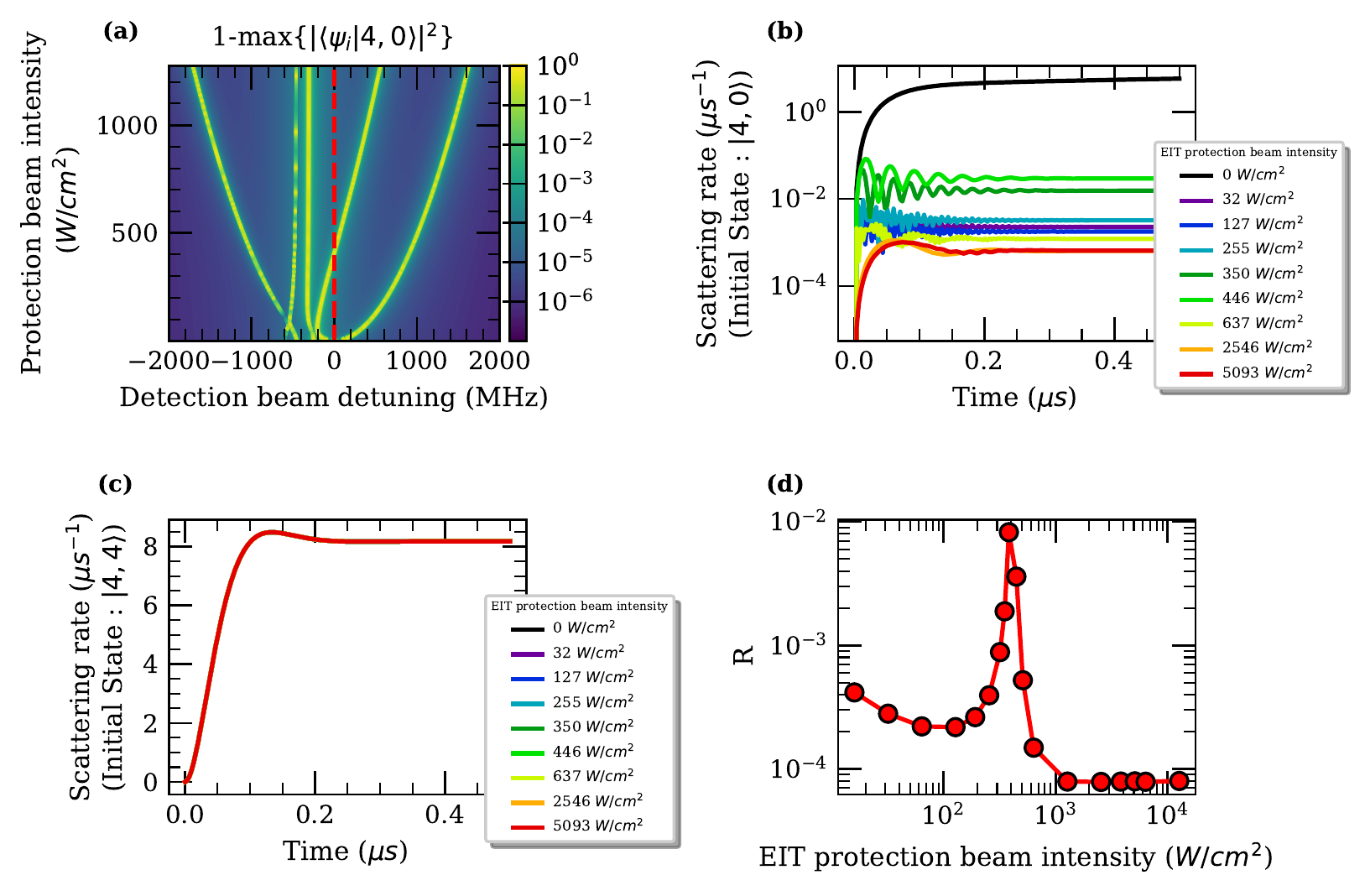}
\caption{\label{6p32result} Scheme 1 simulation results. (a) The non-ground state population of the nominally EIT-protected qubit state as a function of the detection beam detuning and the protection beam intensity. The red dashed line marks where the detection beam is resonant with the cycling transition. (b) The scattering rate in the upper qubit state as a function of time for different intensities. (c) The scattering rate of the population in the stretched state as function of time. All the curves overlap completely since scattering on the cycling transition is unaffected by the EIT beam intensity. (d) The suppression factor for the reabsorption of scattered photons ($R$) as a function of the intensity of the protection beam. The result is proportional to the non-ground state populations shown along the red dashed line in (a). }
\end{figure*}

In order to calculate the performance of the proposed detection scheme, it is necessary to take into account all 64 states shown in Fig.~\ref{scheme1}. Since the light is far-detuned for transitions from $\ket{3,0}$, EIT is not needed for its protection. The main concern is the protection of $\ket{4,0}$. The full Hamiltonian can be diagonalized to obtain the dressed states. Fig.~\hyperref[{6p32result}]{\ref*{6p32result}(a)} shows the extent of mixing of $\ket{4,0}$ with other short-lived states, as a function of both probe beam detuning and protection beam intensity. It is clear that the situation is considerably more complicated than the simple three level EIT shown in Fig.~\hyperref[{3level}]{\ref*{3level}(a)}. In Appendix~\ref{Simplersystems}, by adding one level at a time, we are able to explain the various features. Since the detection transition remains unaffected by the EIT protection, our concern here continues to be the behavior of this graph along the red dotted line.

While the dressed states accurately predict the behaviour for adiabatic preparation, we also solve the master equation to further validate the result, as well as to simulate a situation closer to actual experiments. To this end we have performed the simulation with the population starting in $\ket{4,0}$ and letting the state evolve for up to a few microseconds. The scattering rate from $\ket{4,0}$ as a function of time can be seen in Fig.~\hyperref[{6p32result}]{\ref*{6p32result}(b)}. After an initial transient, within a few hundred nanoseconds the rate settles to a constant on this timescale. The population slowly leaks away from $\ket{4,0}$ to $\ket{3,0}$, $\ket{3,1}$, $\ket{3,2}$, $\ket{4,1}$ and $\ket{4,2}$.

The scattering from the detected state, $\ket{4,4}$, follows the behavior of a simple cycling transition (see Fig.~\hyperref[{6p32result}]{\ref*{6p32result}(c)}), settling to a steady state value after an initial Rabi flop. Combining the results of Figs.~\hyperref[{6p32result}]{\ref*{6p32result}(b)} and \hyperref[{6p32result}]{\ref*{6p32result}(c)}, we find $R$, which is shown in Fig.~\hyperref[{6p32result}]{\ref*{6p32result}(d)} as a function of the protection beam intensity. There are two salient features: the dramatic peak at $~ 400\:\text{W}/\text{cm}^2$ and the saturation of $R$ as the protection beam power is increased. The peak in Fig.~\hyperref[{6p32result}]{\ref*{6p32result}(d)} corresponds to where the red dotted line crosses the sloped line in Fig.~\hyperref[{6p32result}]{\ref*{6p32result}(a)}, where the eigenstate has a large fraction of population in the excited states. The saturation in Fig.~\hyperref[{6p32result}]{\ref*{6p32result}(d)} is also visible in the uniform color in Fig.~\hyperref[{6p32result}]{\ref*{6p32result}(a)} along the upper part of the dotted red line.

The best protection available in Scheme 1 is $R= 8{\times}10^{-5}$. Referring to Eqn.~\ref{Errorin2Dequation} and Eqn.~\ref{Errorin3Dequation}, for a lattice spacing of $5\: \mu \text{m}$, $\gamma=100$, and a total error per detection of $10^{-4}$, the array size can be $62500$ atoms in 2D and only $125$ in 3D. If $10^{-3}$ error per detection is allowed, a 3D geometry would support $157,000$ atoms, while in 2D, the array size would almost certainly never be limited by this source of error. Unlike in Scheme 2, there is no straightforward way to incorporate cooling during the detection.

\section{\label{Scheme2Section}EIT protection Scheme 2}
\begin{figure}
\includegraphics[width=0.45\textwidth]{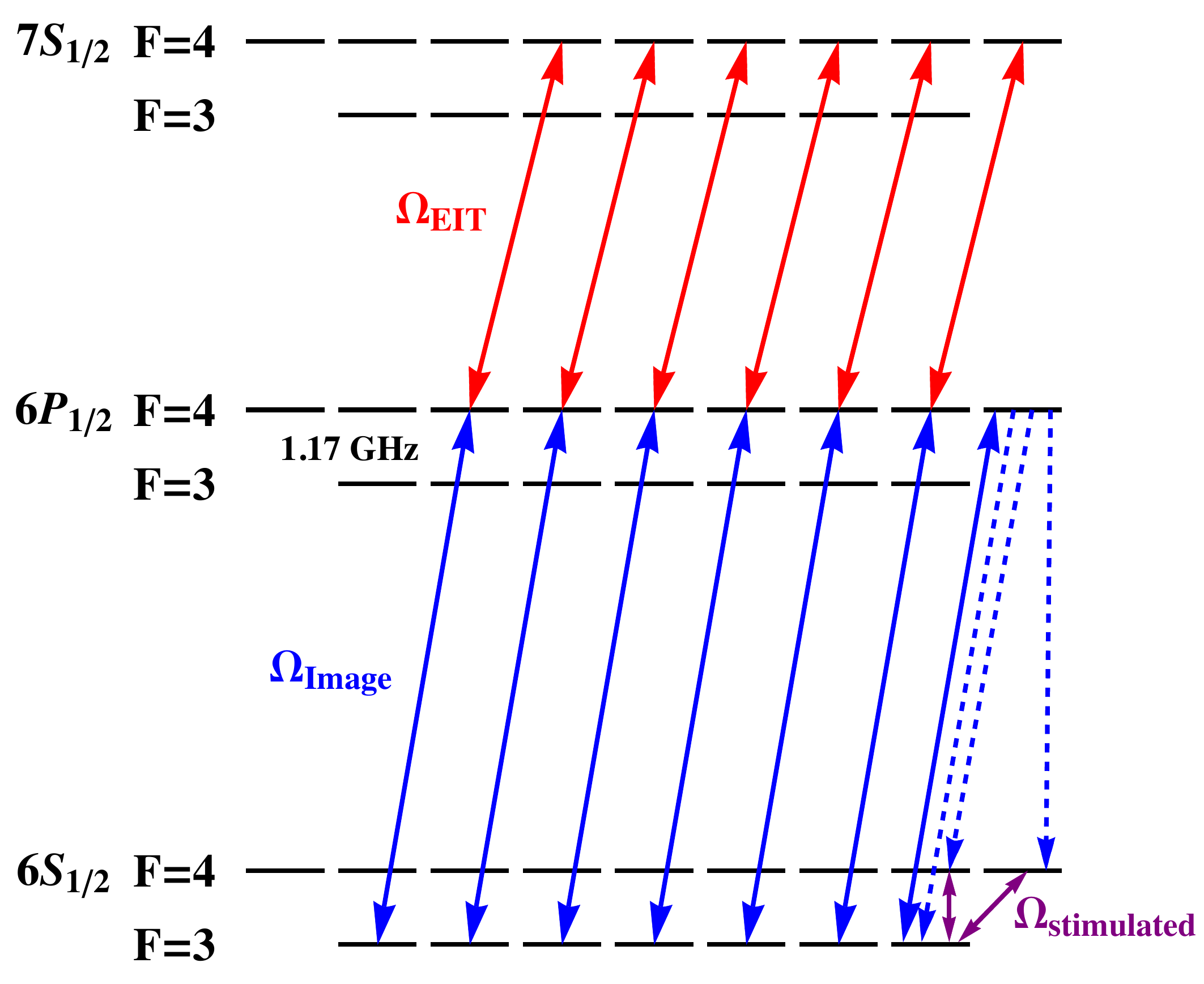}
\caption{\label{Scheme2 fig}EIT-protection Scheme 2. In this case both the imaging and EIT protection light are $\sigma^+$-polarized. The unprotected state is $\ket{3,3}$, with imaging on the open transition $\ket{3,3}\longrightarrow\ket{4',4'}$. A set of stimulated pulses can be used to get the population out of the dark $\ket{4,3}$ and $\ket{4,4}$ states in order to continue imaging the atoms. These pulses can additionally drive sideband transitions to simultaneously cool the atoms.}
\end{figure}

Scheme 2, shown in Fig.~\ref{Scheme2 fig}, is an example of imaging using an open transition. Here the $\ket{3,0}\longrightarrow\ket{4',1'}$ is protected. We take $6P_{1/2}$ to be the intermediate state because the larger hyperfine splitting and the reduced number of hyperfine levels compared to $6P_{3/2}$ greatly reduce multilevel effects. We note, however, that $6P_{3/2}$ could also be used, albeit with worse performance. As can be seen in Fig.~\ref{Scheme2 fig}, the imaging transition $\ket{3,3}\longrightarrow\ket{4',4'}$ is not closed, so that for a long enough imaging time, all of the population will end up in $\ket{4,3}$ and $\ket{4,4}$, which are dark to the imaging light. 

\subsection{\label{scheme2 imaging}Imaging on an open transition}

To repeatedly scatter photons from $\ket{3,3}$, it is necessary to recover the population from $\ket{4,3}$ and $\ket{4,4}$. The idea is to image the atoms for some time $\tau$, losing the population to $\ket{4,3}$ and $\ket{4,4}$. Then stimulated pulses can be used to shuffle the populations among the three sublevels, before proceeding with the imaging again. In particular, the sequence can be as follows: 

\begin{enumerate}
    \item $\sigma^+$-polarized detection beam is turned on for some time $\tau$ and scattering $N$ photons. Since both the signal and total error are proportional to each other, the protection is independent of $\tau$. Therefore the pulse can be made long enough to significantly deplete the population in $\ket{3,3}$.
    \item A pulse is applied to exchange the populations between $\ket{4,3}$ and $\ket{3,3}$. The imaging light is turned on for $\tau$, again depleting the population in $\ket{3,3}$. This can be repeated until the population is overwhelmingly in $\ket{4,4}$.
    \item Another pulse is applied to exchange the populations between $\ket{4,4}$ and $\ket{3,3}$.
    \item The entire sequence is repeated.
\end{enumerate}

This sequence of steps leads to a closed-loop imaging, where the atoms are effectively pumped to $\ket{4,4}$ and the imaging can be restarted, so that the atoms can be efficiently detected. The key advantage of Scheme 2 is that the reshuffling pulses can be used to drive sideband transitions to cool the atoms. Because cooling is compromised when there are initially atoms in the final state, it is best to significantly empty $\ket{3,3}$. To steadily scatter photons, the steady state population during detection must be clearly above the vibrational ground state. But once enough photons have been detected, a final cooling sequence can focus on transferring only vibrationally excited atoms back to $\ket{3,3}$, ultimately leaving most of the atoms in the vibrational ground state of $\ket{4,4}$. Having this efficient way of cooling and ending with the atoms in a well-defined state after detection allows atoms to be reset for further quantum computation after the imaging.

\begin{figure}
\centering
\includegraphics{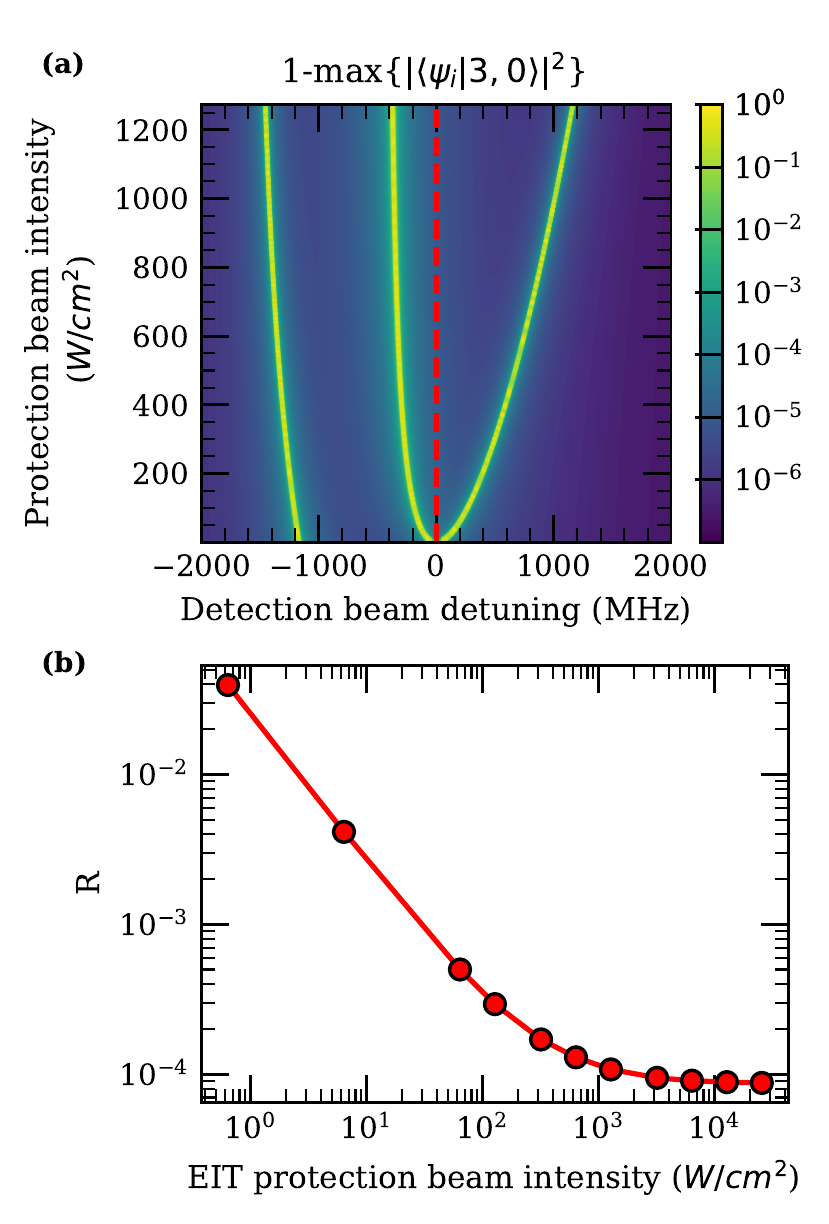}
\caption{\label{Scheme 2 results} Scheme 2 results. (a) The non-ground state population of the EIT-protected qubit state $\ket{3,0}$ as a function of the detection beam detuning and the protection beam intensity. The red dashed line marks where the detection beam is resonant with the open transition for imaging. (b) The suppression factor for the reabsorption of scattered photons ($R$) as a function of intensity of the protection beam. }
\end{figure}

We follow a similar methodology to the previous section to characterize the EIT protection. As expected, the population in $\ket{3,3}$ settles to an exponential decay after an initial Rabi flop, while the $\ket{3,0}$ can be characterized by a linear decay. $\ket{4,4}$ and $\ket{4,3}$ are both far detuned from the imaging light and are therefore orders of magnitude better protected than $\ket{3,0}$. This means that any population that ends in either of these states remains there and any error caused by them can be safely neglected. Using this information, we proceed to characterize $R$ as shown in Fig.~\ref{Scheme 2 results}. The similar performance to Scheme 1 and the atom reusability make this scheme the preferred choice despite it being more technically and conceptually complicated.

Having an open transition for imaging allows for the possibility of using a forbidden transition that offers better protection of the qubit states from direct or stray light. The idea is to, for example, protect through the forbidden $\ket{4,0}\longrightarrow\ket{4',0'}$ and image on the open $\ket{4,4}\longrightarrow\ket{4',4'}$ transition. The zero matrix element means that the qubits will be better protected from the $\pi$-polarized detection beam. However, it doesn’t offer significantly better protection for randomly polarized rescattered light, which we are mainly concerned about here. In practical cases where direct/stray light is a major problem, this protection could be beneficial. It does come at the expense of a more elaborate detection procedure and an extra source of protection errors. For a longer discussion about this possibility, which we call Scheme 3, see Appendix~\ref{forbidden-transition}.

\section{\label{PolarizationImperfectionsSection}Polarization Imperfections}

In any real experiment, residual polarization errors will compromise the proposed detection. Since Scheme 2 seems like the best choice, we will analyze this case. There are two different types of polarization errors that can affect the detection scheme, imperfect protection beam polarization and imperfect detection beam polarization.

\begin{figure}
\includegraphics[width=0.44\textwidth]{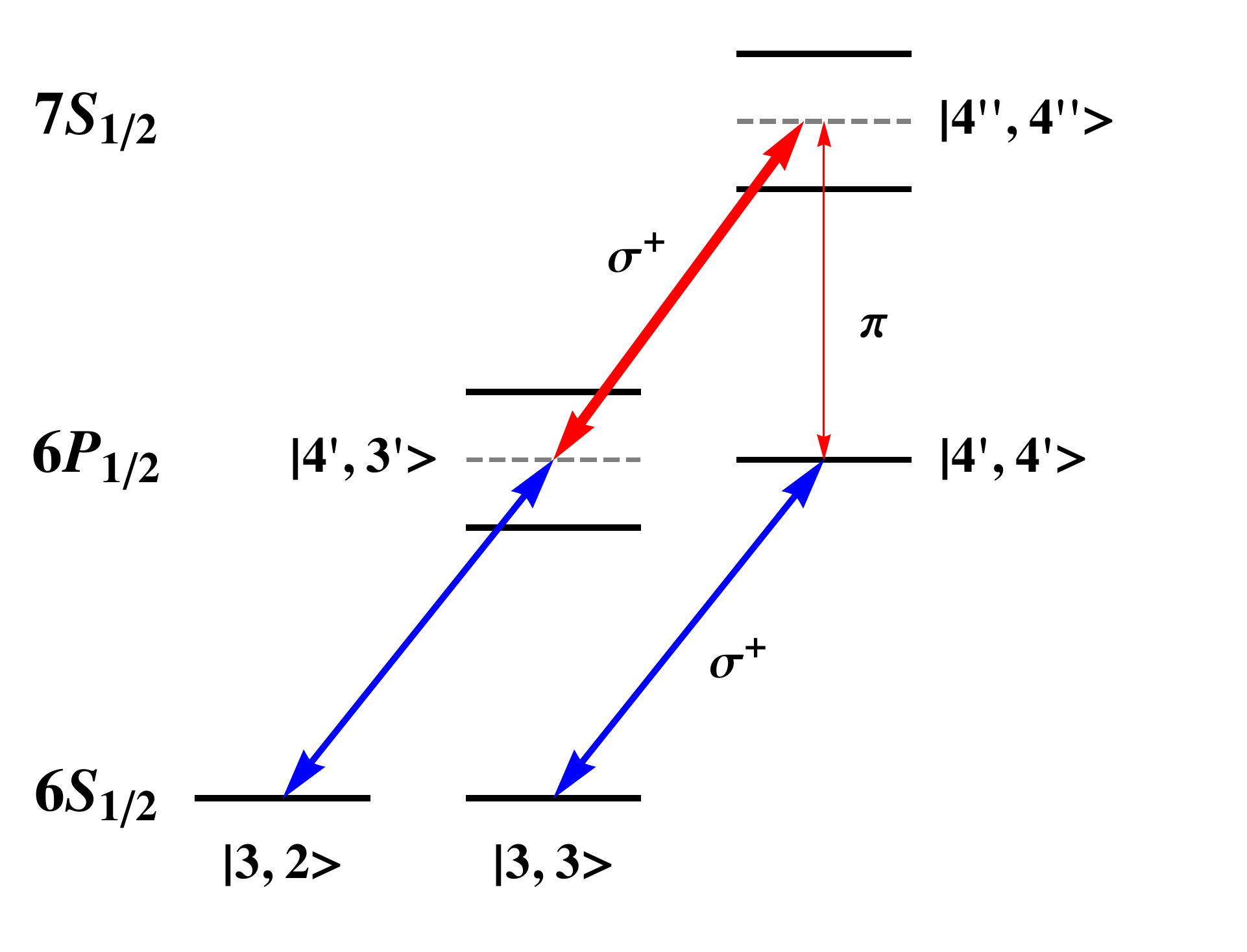}
\caption{\label{Scheme2Err1}Scheme 2 possible error caused by a small amount of the wrong $\pi$-polarization in the protection beam. The $\pi$-polarization (denoted by the thin red arrow), which would otherwise resonantly couple $\ket{4',4'}$ to a bare $\ket{4'',4''}$ (upper right dashed grey line), is off-resonant due to the strong $\sigma^+$ (denoted by the thick red arrow), which Rabi-splits $\ket{4'',4''}$ (upper right solid black lines).}
\end{figure}

For the first type of error (see Fig.~\ref{Scheme2 fig}), if some $\sigma^-$ or $\pi$-polarized light is added, state $\ket{3,0}$ remains protected. However, $\pi$ or $\sigma^-$-polarized light could potentially cause $\ket{3,3}$ to become EIT protected, interfering with detection. A careful examination reveals that this is will not be a problem. Consider Fig.~\ref{Scheme2Err1}, where there is a little bit of $\pi$-polarization. $\ket{4',4'}$ cannot be coupled to $\ket{4'',4''}$ because $\ket{4'',4''}$ is already very strongly coupled to $\ket{4',3'}$. The $\pi$-polarized light is prevented from accidentally EIT protecting the detection transition because it is itself EIT-protected on its own transition. 

\begin{figure}
\centering
\includegraphics{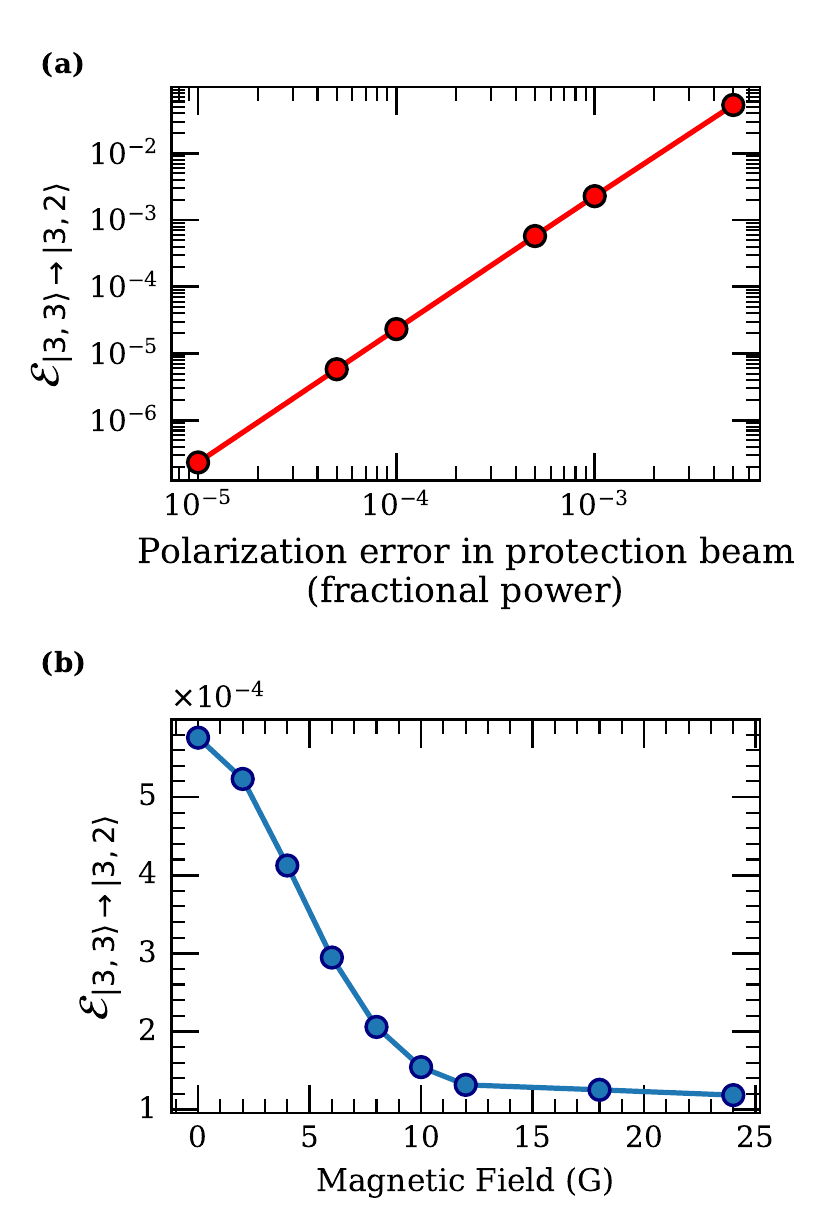}
\caption{\label{Scheme2 Polarization Error}Scheme 2 polarization error in EIT protection beam. (a) A varying fractional power is applied in the wrong $\pi$-polarization in the protection beams (with no magnetic field). The main source of error is population transfer $\ket{3,3}\longrightarrow\ket{3,2}$ through a four-photon Raman transition. (b) Keeping the fractional power at $5\times10^{-4}$, a magnetic field of varying size is applied in order to reduce the error.}
\end{figure}

The main source of error comes from the fact there is a possible four-photon Raman transition $\ket{3,3}\longrightarrow\ket{3,2}$. However, two of these transitions ($\ket{4',4'}\longrightarrow\ket{4'',4''}$ and $\ket{4',3'}\longrightarrow\ket{3,2}$) are single-photon off-resonant and involve weak beams. As can be seen in Fig.~\hyperref[{Scheme2 Polarization Error}]{\ref*{Scheme2 Polarization Error}(a)}, this $4^{th}$ order process causes only a small error.

Loss to $\ket{3,2}$ can be suppressed by using a magnetic field to make this four-photon transition off-resonant. In order to characterize this source of error we have computed the error as a function of fractional power in the wrong $\pi$-polarization and in different magnetic fields, as shown in Fig.~\ref{Scheme2 Polarization Error}. For an experimentally achievable fractional power with the wrong polarization of $\sim10^{-4}$, the error can be kept below $10^{-4}$ even with no additional magnetic field (again assuming 100 photons for imaging). However, if the fractional power is significantly higher, a magnetic field can be used to further suppress this source of error.

Imperfect detection beam polarization does not cause unwanted transitions to the intermediate state because all those states are EIT-protected, nor does it lead to any Raman transitions to the adjacent level. For the previous error, although the $\ket{3,2}\longrightarrow\ket{4',3'}$ is EIT protected, the two-photon $\ket{3,2}\longrightarrow\ket{4'',4''}$ transition can still happen, which enables the four-photon process. Here, on the other hand, the $\ket{3,2}\longrightarrow\ket{4',3'}$ EIT light directly suppresses the error. We see no such transitions in our simulation with up to $5{\times}10^{-3}$ of the wrong polarization.

From this discussion, it is clear that polarization imperfections only moderately impact the EIT scheme. Residual effects can be mitigated by controlling adjustable experimental parameters like magnetic fields.

\section{\label{Discussion}Conclusion}
We have presented a way to measure the states of selected qubits in an atom array while preserving the surrounding quantum information. Such a capability is needed for quantum error correction and one-way quantum computation. To a first approximation, one can detect selected atoms by only shining detection light at those atoms. Our approach, EIT-protecting the unselected atoms, addresses the problem that they can rescatter detection light, as well as the limitation that imperfect beam quality or scattering from surfaces can also lead to to some direct illumination of what are supposed to be the spectator atoms. One type of scheme we consider allows for cooling while detecting, so that the detected atoms can be reused.

In summary, assuming that 100 scattered photons are needed for detection, we find that for 2D arrays, EIT protection is sufficient to allow for having $>$60,000 atoms while maintaining an error per measurement below $10^{-4}$. Only about 125 atoms can be used in a 3D array before that measurement threshold is exceeded, at least for measurement of the central atom. Relaxing the error threshold to $10^{-3}$ would allow for $>$275,000 atoms in the array. Without relaxing the error threshold, a similar number of atoms can be used if the required number of scattered photons needed for detection was reduced to 10, for instance by phase contrast measurement techniques. 

There are other possible ways to selectively measure atoms, like using a second species or moving atoms to a more distant location for measurement. Even in the same apparatus, one can imagine a role for each of these approaches.

\begin{acknowledgments}
This work was supported by AFOSR grant FA9550-20-1-0110 and NSF grant PHY-1820849.

F.G. and A.K. contributed equally to this work.

\end{acknowledgments}

\appendix

\section{\label{Mapping}Coherent mapping out of the Qubit States}

\begin{figure}
\includegraphics[width=0.45\textwidth]{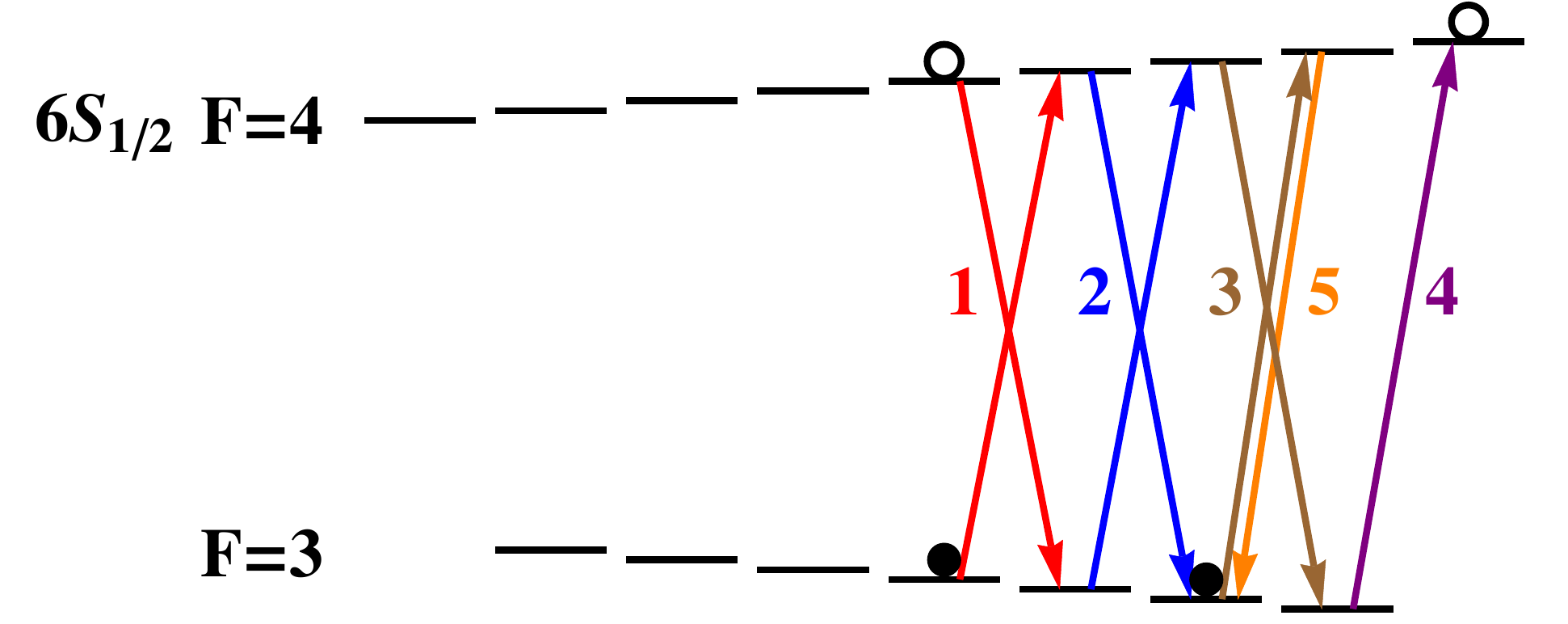}
\caption{\label{gnd_tranfer}Selective transfer of an atom out of the qubit basis. The first step (1) in the sequence requires site-specificity, while the remaining steps do not, as they are not resonant with atoms in the qubit states.}
\end{figure}

For Cs, where the clock states ($\ket{3,0}$ and $\ket{4,0}$) are a convenient qubit basis, a possible mapping proceeds as in Fig.~\ref{gnd_tranfer}. The first step, to $\ket{4,1}$ and $\ket{3,1}$, is the only one that requires spatial selectivity (sequentially or in parallel). In 3D this can be accomplished as in \cite{Wang2015_3Daddr}, using microwave adiabatic fast passage (AFP) pulses and a pair of crossed ac-Stark shifting addressing beams. In 1D or 2D, it can be accomplished either with microwave pulses and a single addressing beam or with two-photon Raman transitions. The frequency selectivity afforded by Zeeman shifts will then allow all subsequent transfers to be accomplished using only microwaves. Two pulses will map the states to $\ket{4,3}$ and $\ket{3,3}$. Then one pulse will map $\ket{3,3}\longrightarrow\ket{4,4}$. A final pulse will map $\ket{4,3}\longrightarrow\ket{3,2}$, where it will remain protected while the first state is measured. After the measurement of the first state, the two states will be swapped to measure the second one. Although somewhat redundant, the measurement of the second state will reveal if atoms have been lost. After both measurements, the atom, now in a known state, can be returned to a state in the qubit basis by essentially reversing Fig.~\ref{gnd_tranfer}. The returning sequence will again only necessitate one single spatially selective transfer per atom.

\section{\label{Simplersystems}Eigenstate calculations for simplified systems}

We numerically calculate the eigenstates of some simpler Hamiltonians to get insight into the behaviour of EIT protection. The essential features we would like to explain are the saturation of the EIT protection, as well as the scattering resonance that is observed in Fig.~\hyperref[{6p32result}]{\ref*{6p32result}(d)}, which is why for these calculation we use $6P_{3/2}$ as the intermediate state. We calculate the eigenstates ($\psi_i$) for several Hamiltonians with increasing complexity and plot the quantity $1-max\{|\langle\psi_i|4,0\rangle|^2\}$, i.e., the total intermediate and excited state populations for the eigenstate that has the highest projection on the ground state, while varying the detection beam detuning and the protection beam intensity. For adiabatic preparation, this quantity is directly proportional to the scattering rate from the ground state.

\begin{figure}
\includegraphics[width=0.45\textwidth]{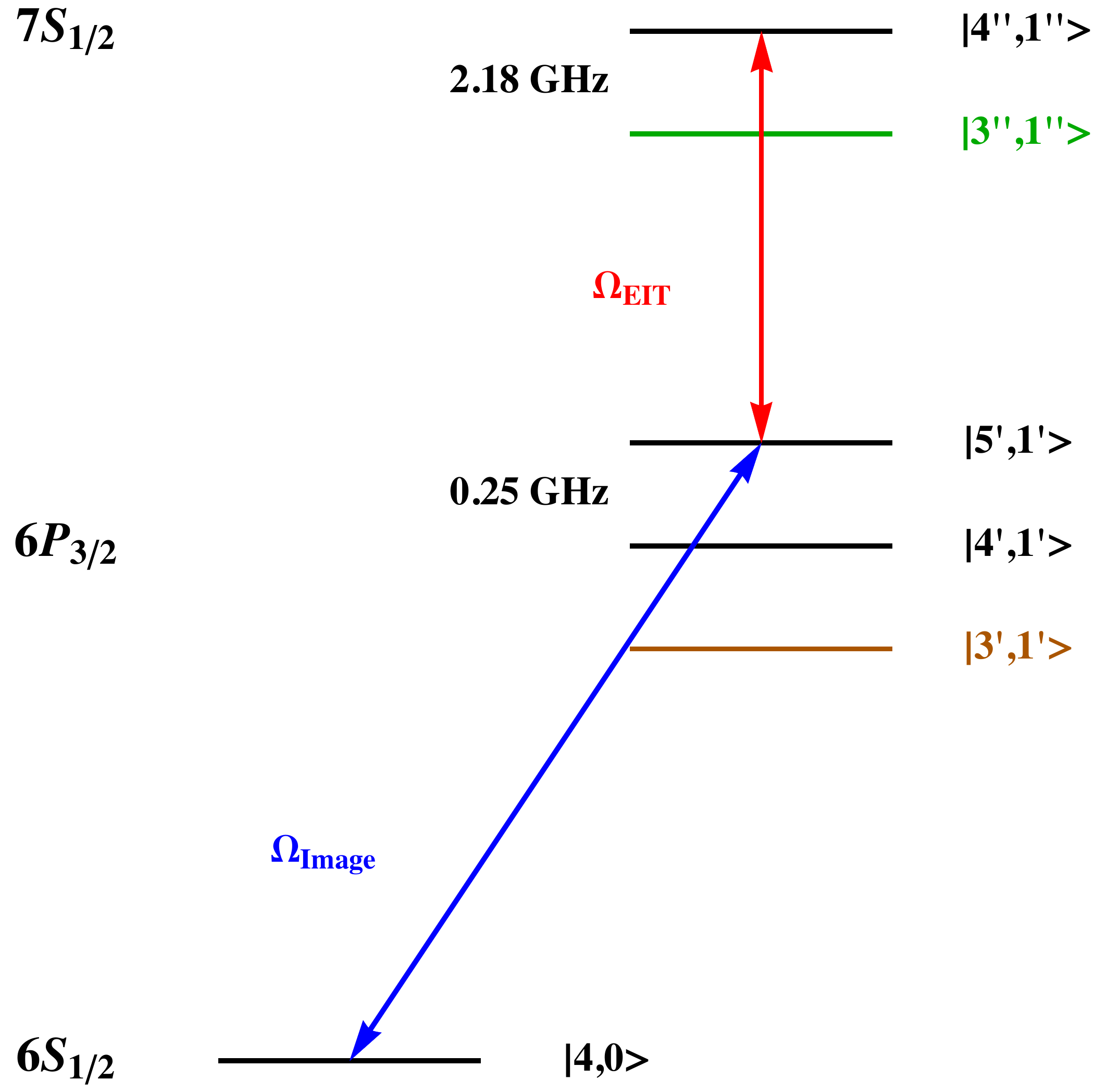}
\caption{\label{simplelevels}Level diagram for simplified EIT-protection systems discussed in this appendix. The basic four levels in black, $\ket{4,0}$, $\ket{4',1'}$, $\ket{5',1'}$ and $\ket{4'',1''}$, are discussed in Fig.~\ref{fourlevel}. Fig.~\ref{fivelevel} adds in the green level, $\ket{3'',1''}$. Finally, Fig.~\ref{sixlevel} adds in another level in brown, $\ket{3',1'}$. }
\end{figure}

\begin{figure}
\includegraphics[width=0.45\textwidth]{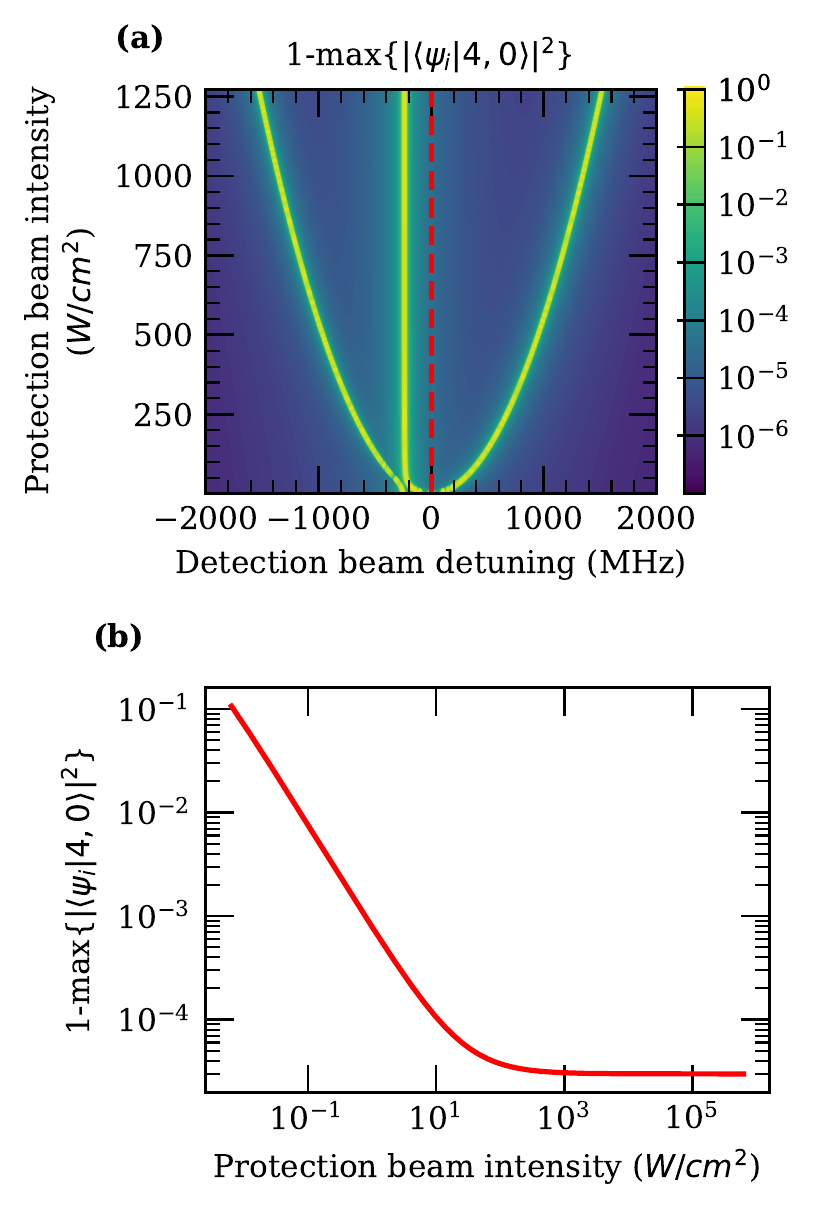}
\caption{\label{fourlevel}EIT-protection of the 4 level-system with states $\ket{4,0}$, $\ket{4',1'}$, $\ket{5',1'}$ and $\ket{4'',1''}$. (a) Compared to the results in Fig.~\ref{3level}, the new bright state has constant detuning from the resonant detection beam (red dashed line) at high protection beam intensities. (b) The addition of the new level causes the saturation of $R$.}
\end{figure}

\begin{figure}[htb]
\includegraphics[width=0.45\textwidth]{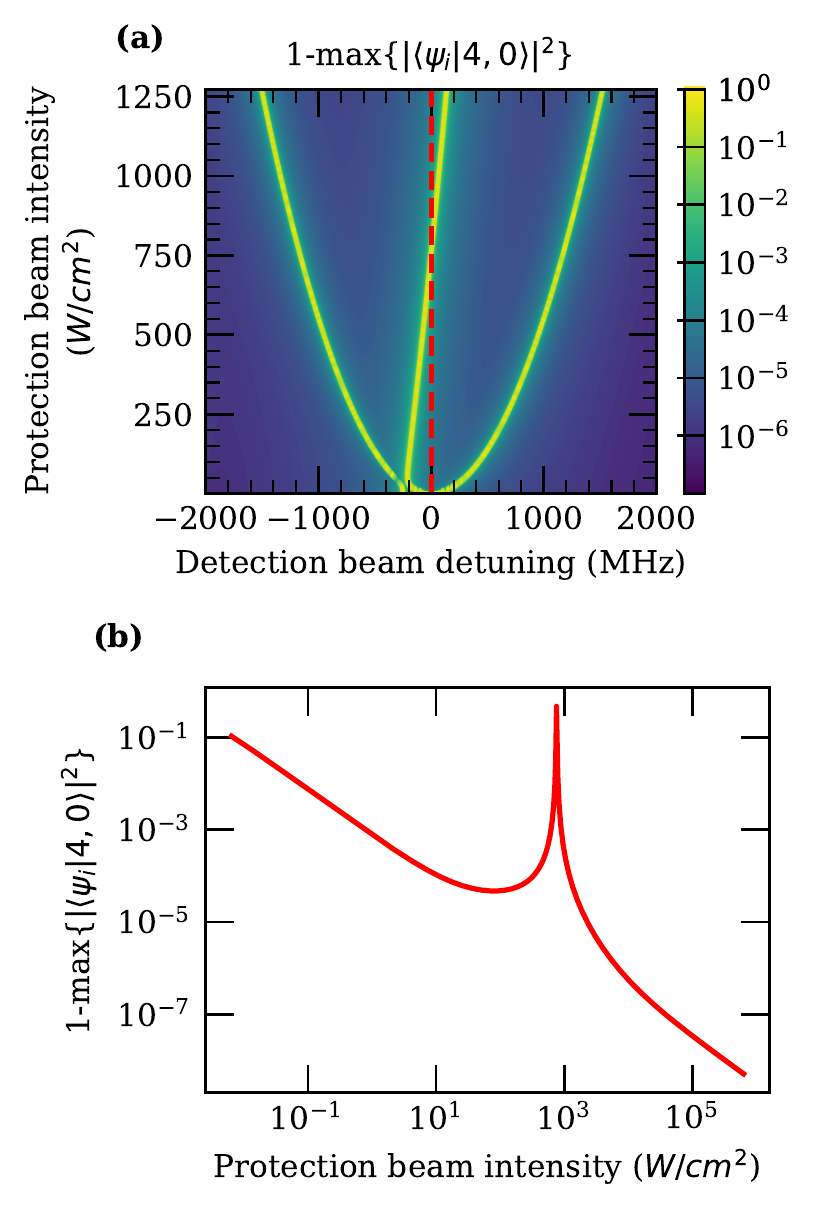}
\caption{\label{fivelevel} EIT-protection of the five level system with the same states as the previous four level system plus $\ket{3'',1''}$. (a) The new bright state compared to Fig.~\ref{3level} has a varying detuning in this case, with the positive slope causing it to cross 0 detuning (red dashed line) at around $800\:\text{W}/\text{cm}^2$ . (b) The addition of the new level causes the peak of $R$ at the crossing point in (a).}
\end{figure}

We start with a four level system containing only the states $\ket{4,0},\,\ket{4',1'},\,\ket{5',1'},\,\text{and} \,\ket{4'',1''}$, and protection and detection beam configurations shown by the black levels in Fig.~\ref{simplelevels}. Fig.~\ref{fourlevel} shows the results of this calculation. As the protection beam intensity is increased, the usual 3-level EIT splitting is observed until the EIT window gets close to twice the hyperfine splitting of the intermediate state (Fig.~\hyperref[{fourlevel}]{\ref*{fourlevel}(a)}). This is also evident in the initial linear decay with a slope of -1 in the log-log plot in Fig.~\hyperref[{fourlevel}]{\ref*{fourlevel}(b)} at zero detection beam detuning. The linear decay is followed by a saturation of the excited state population due to the existence of a constant `bright' eigenstate at a fixed detuning from the detection beam. 

One can get insight into this behavior by considering the case where the two intermediate states are degenerate, in which case the two intermediate states and the excited state themselves form a lambda type three level system. The `dark' state of this three level system is also given by Eqn.~\ref{eqn1}, where the mixing angle, $\theta$, is simply the ratio of the matrix elements of the two intermediate states to the excited state (because they are coupled by the same field). Since that ratio is fixed, this `dark' state is also fixed. But in general this `dark' state can couple to the ground state in the presence of a detection field. This reasoning can also be extended to the case of a nonzero hyperfine splitting, $\delta$. In the limit $\Omega_{EIT} >>\delta$, such a fixed state still exists, but is simply detuned from the probe beam.

Next we add the state $\ket{3'',1''}$ to the previous Hamiltonian (see the black and green levels in Fig.~\ref{simplelevels}), with the results shown in Fig.~\ref{fivelevel}. Fig.~\hyperref[{fivelevel}]{\ref*{fivelevel}(a)} shows that the previously constant `bright' state will be shifted into resonance as the protection beam power is increased. The shift is due to the competing (different sign of detunings) ac-Stark shifts of the $\ket{4',1'}$ state due to the coupling with the $\ket{3'',1''}$ and $\ket{4'',1''}$ states. Although the protection beam is much further detuned from the $\ket{4',1'}$ to $\ket{3'',1''}$ transition (1936 MHz) than the $\ket{4',1'}\longrightarrow\ket{4'',1''}$ transition (-250 MHz), since the matrix element for the former is larger, it eventually overpowers the Stark shift due to the latter. Since the detuning is no longer fixed, the linear decrease in error with protection beam intensity is recovered after the peak. It is intriguing that one can get this desired linear error decrease even for a five level system. The lack of $\ket{5',1'}$ to $\ket{3'',1''}$ coupling seems to be critical. Unfortunately, the real Cs atom has another relevant level. It would be possible to achieve this situation for a spin-1/2 atom with a $P-D'-P''$ ladder, which is not available in typical cold atom systems.

\begin{figure}[htb]
\includegraphics[width=0.45\textwidth]{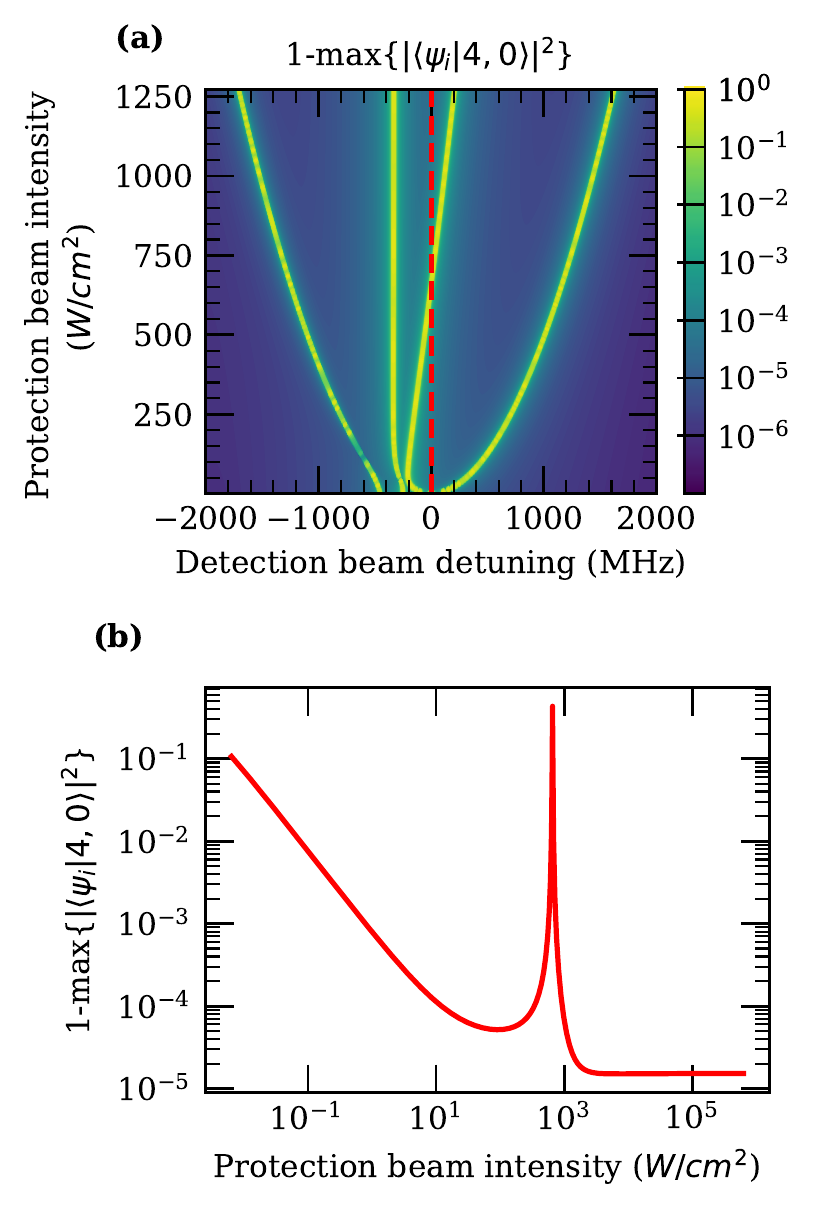}
\caption{\label{sixlevel}EIT-protection of the six level system with the same states as the previous five level system plus $\ket{3',1'}$. (a) There are two new bright states compared to Fig.~\ref{3level}. One of these two states has a constant detuning at high protection beam intensities, while the other has a positive slope and crosses 0 detuning (red dashed line) at around $800\:\text{W}/\text{cm}^2$. (b) The combination of these two new bright states cause the peak of $R$ at the crossing point in (a) and the saturation at high protection beam intensities.}
\end{figure}

To explain the saturation in the full system the final relevant level, $\ket{3',1'}$ needs to be added to the Hamiltonian (see all levels in Fig.~\ref{simplelevels}). Fig.~\ref{sixlevel} shows that the fixed `bright' eigenstate is recovered, which yields a combination of the two features previously explained.

\section{\label{forbidden-transition}Scheme 3}

\begin{figure}
\includegraphics[width=0.45\textwidth]{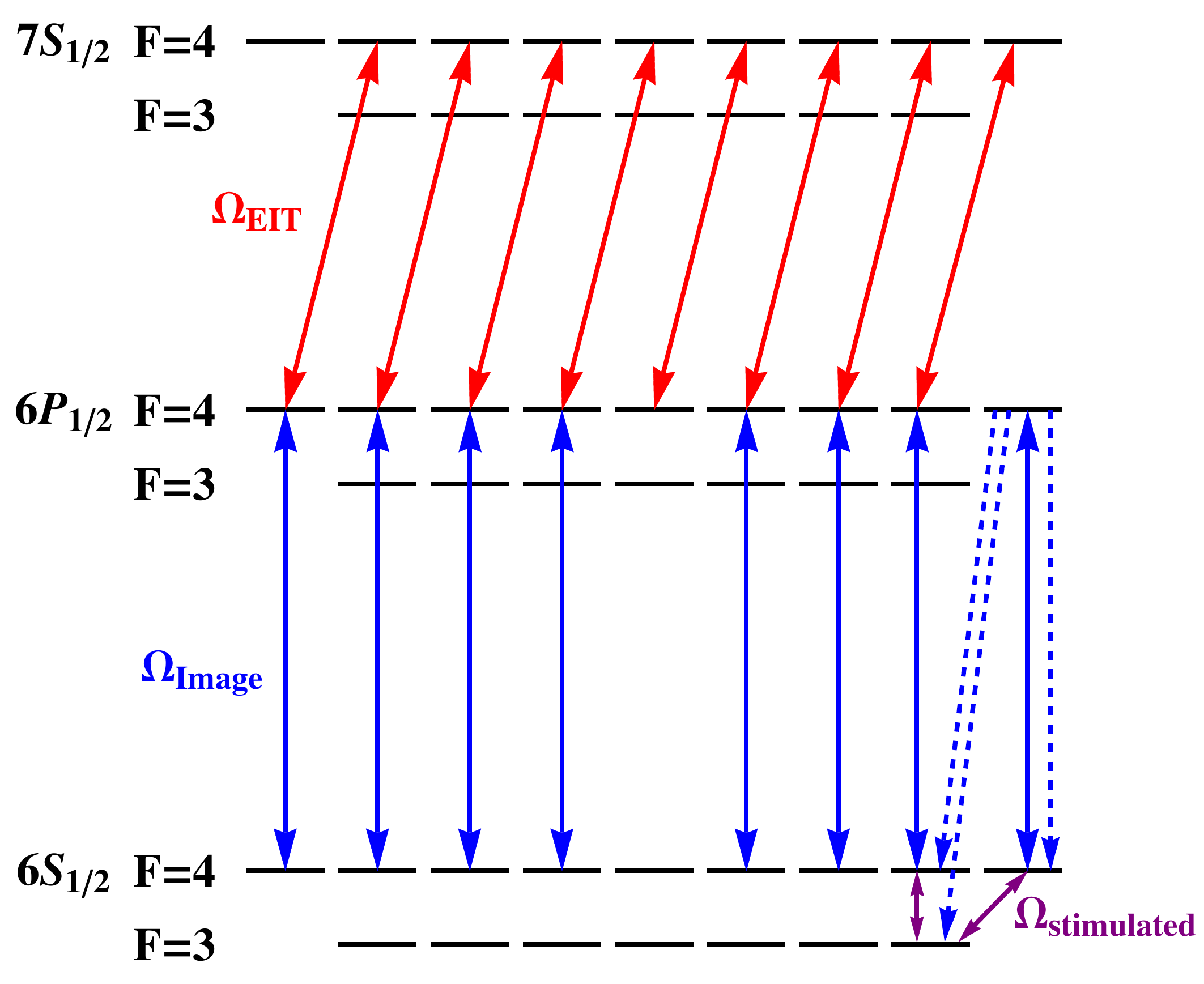}
\caption{\label{forbidden-transition-EIT}EIT-protection using a forbidden transition. The qubit state $\ket{4,0}$ is protected from the $\pi$-polarized imaging light by selection rules. All other F=4 states are EIT protected except $\ket{4,4}$.}
\end{figure}

As mentioned at the end of Section~\ref{Scheme2Section}, there exists the possibility to further suppress errors from direct light through a variation of Scheme 2, which we will call Scheme 3, that takes advantage of a forbidden transition. For instance, when the intermediate state is $F'=4'$ and $\pi$-polarized detection light is used, the $\ket{4,0}\longrightarrow\ket{4',0'}$ transition is forbidden. EIT protection of the $\ket{4,0}$ state is only needed to protect against excitation due to imperfectly polarized light, such as rescattered light. EIT also protects other $m_F$ levels, especially the $\ket{4,3}$ state, which gets populated because the imaging occurs on an open transition. The best choice for the EIT protection transition is $6P_{1/2},\,F'=4'$ to $7S_{1/2},\,F''=4''$ with a $\sigma^+$-polarized beam, as shown in Fig.~\ref{forbidden-transition-EIT}. Another possible choice would be to use the $6P_{1/2},\,F'=4'$ to $7S_{1/2},\,F''=3''$ transition with a $\pi$-polarized beam, but in that case, off-resonant scattering of the imaged state from the $7S_{1/2},\,F''=4''$ state compromises the detection fidelity.

\begin{figure}
\includegraphics[width=0.45\textwidth]{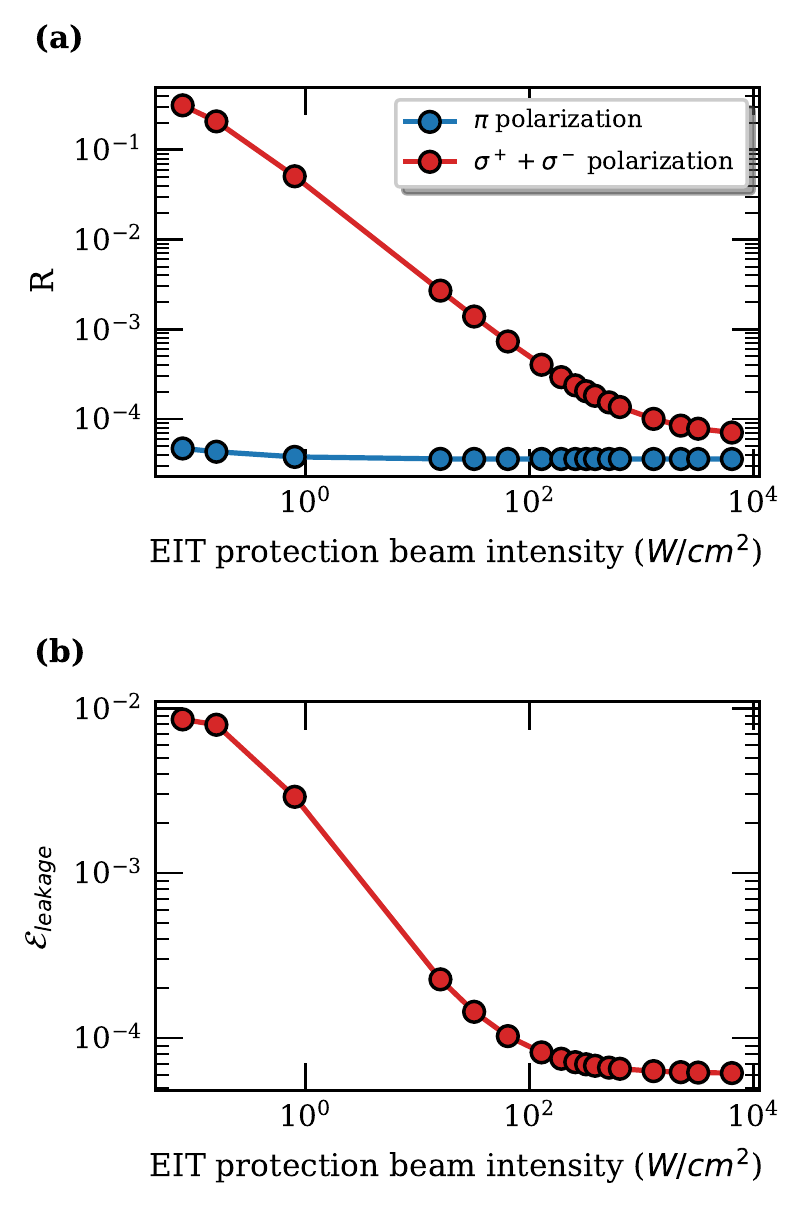}
\caption{\label{forbidden-transition-results} Scheme 3 protection and leakage error results. (a) The suppression factor for the reabsorption of scattered photons ($R$) as a function of intensity of the protection beam, for $\ket{4,0}$ with $\pi$-polarized light (in red, corresponding to the level structure in Fig.~\hyperref[{forbidden-transition-dressed}]{\ref*{forbidden-transition-dressed}(a)}) and $\sigma^++\sigma^-$ light (in blue). In general, the protection will be some linear combination of these results depending on the polarization of the scattered light. (b) The leakage error from $\ket{4,3}$ (corresponding to Fig.~\hyperref[{forbidden-transition-dressed}]{\ref*{forbidden-transition-dressed}(b)}) as a function of protection beam intensity.}
\end{figure}

For the detected atoms, as light is scattered on the $\ket{4,4}\longrightarrow\ket{4',4'}$ transition, population accumulates in $\ket{4,3}$ and $\ket{3,3}$. While $\ket{3,3}$ is extremely well protected because it is far off resonant, imperfect EIT protection of $\ket{4,3}$ means that population can find its way to $m_F < 3$ levels, a loss from the qubit basis. Therefore we need to characterize not only the scattering rate from $\ket{4,0}$ for the spectator atoms, but also a new source of error that comes from population loss from $\ket{4,3}$ for the detected atoms, which we call $\mathcal{E}_{leakage}$. We follow a similar methodology as in the main paper to characterize the EIT protection. However, for this scheme, the exact calculation of this error rate depends on the imaging time $\tau$, with different optima for minimizing the errors for detected and spectator atoms. Here we show the result assuming a $\tau$ that reasonably minimizes both errors. Also, for $\ket{4,0}$ we separate the protection between $\pi$-polarized and any other polarization in order to show the enhanced protection for direct/stray light. The results can be seen in Fig.~\ref{forbidden-transition-results}.

\begin{figure}
\centering
\includegraphics[width=0.45\textwidth]{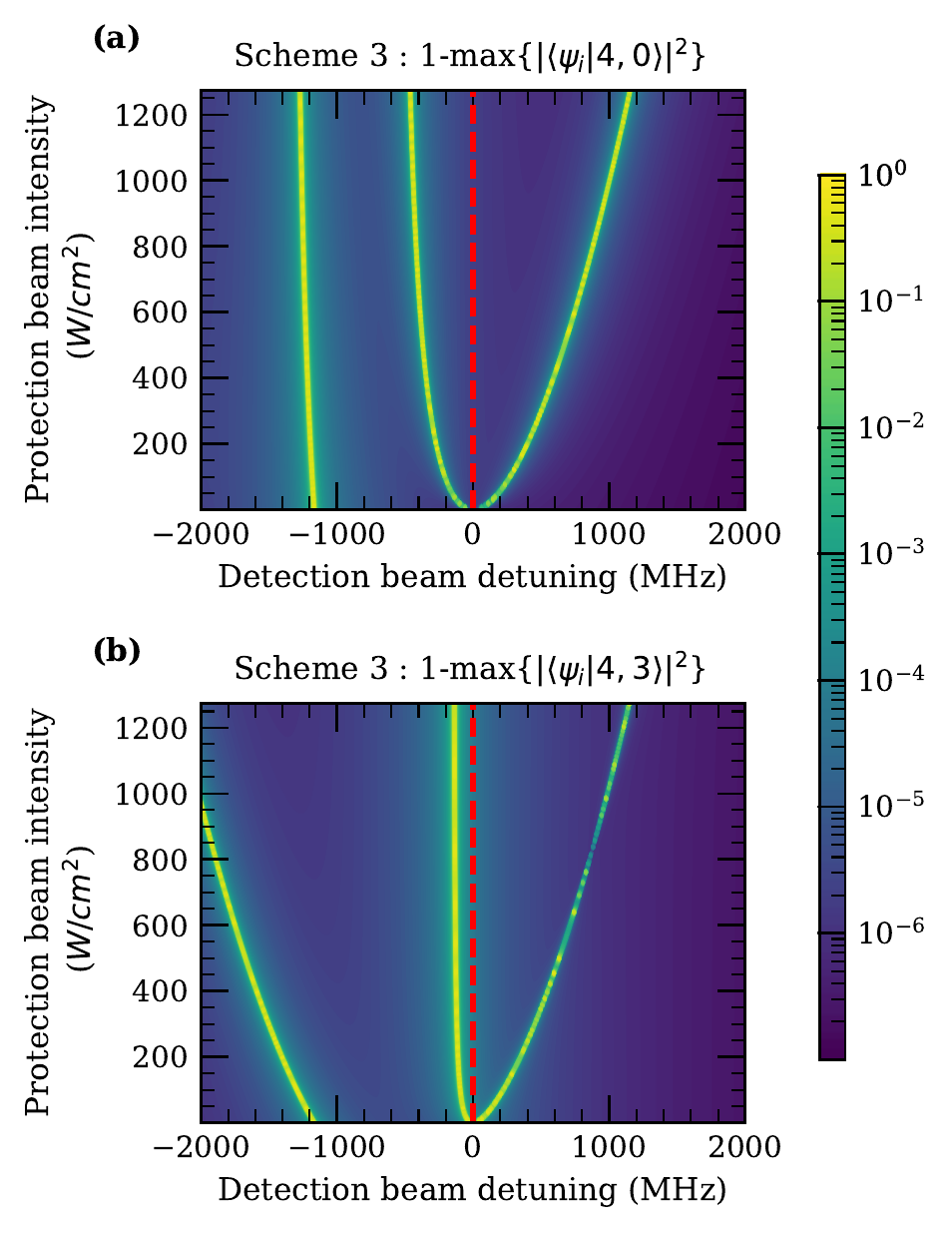}
\caption{\label{forbidden-transition-dressed} Scheme 3 dressed states results. The non-ground state population of the nominally EIT-protected states $\ket{4,0}$ (a) and $\ket{4,3}$ (b) as a function of the detection beam detuning and the protection beam intensity. The red dashed line marks where the detection beam is resonant with the cycling transition.}
\end{figure}

Assuming perfect $\pi$-polarization, the $\ket{4,0}$ state does not need EIT protection, although there is a small amount of off-resonant scattering from $\ket{3',0'}$. Increasing the EIT power initially enhances scattering from the two-photon channel $\ket{4,0}\longrightarrow\ket{3',0'}\longrightarrow\ket{4'',1''}$. Ultimately, EIT protection kicks in for that transition, and the same kind of saturated EIT protection occurs as in Scheme 2, but at a roughly 2.5 times lower error level.

The loss for other polarizations and for $\ket{4,3}$ are suppressed by EIT protection (red circles) and decrease with increasing EIT intensity. Similar to the Schemes 1 and 2, the EIT suppression eventually saturates. Insight into this saturation can be obtained from the structure of the dressed states shown in Fig.~\ref{forbidden-transition-dressed}. In both Figs.~\hyperref[{forbidden-transition-dressed}]{\ref*{forbidden-transition-dressed}(a)} and \hyperref[{forbidden-transition-dressed}]{\ref*{forbidden-transition-dressed}(b)}, the energy of the nearest dressed state with significant excited state populations reaches a fairly constant value at high protection beam intensity, which can be seen by the asymptotically vertical yellow lines at slightly negative detuning. Therefore, the detuning of the resonant detection beam from those states saturates, saturating the EIT protection. 

Since the EIT protection is not perfect for $\ket{4,3}$, there is a cost for keeping the imaging light on for too long. It is best to image for only one period $\tau$, apply a pulse to exchange the populations between $\ket{4,4}$ and $\ket{3,3}$, and then one to exchange the populations between $\ket{3,3}$ and $\ket{4,3}$, and then re-image.

This sequence of steps eventually leads to a pseudo-steady state, where the initial population in each of the relevant sublevels after an iteration is roughly the same as the previous one. The steady state depends on $\tau$, the exact optimization of which requires consideration of the array geometry and the fraction of qubits measured. It should also be noted that it may be possible to recover the population loss associated with $\mathcal{E}_{leakage}$ by using a repumping beam and microwave pulses, so that this error could be mitigated. This is may also be true for the error in Sec.~\ref{PolarizationImperfectionsSection}.


\bibliography{EIT_Measurement_References.bib}

\end{document}